# Phenomenological analysis of densification mechanism during spark plasma sintering of MgAl$_2$O$_4$

Guillaume Bernard-Granger[a], Nassira Benameur[a], Ahmed Addad[b], Mats Nygren[c], Christian Guizard[a] and Sylvain Deville[a]

[a] Laboratoire de Synthèse et Fonctionnalisation des Céramiques
UMR CNRS/Saint-Gobain 3080, Saint-Gobain C.R.E.E.
84306 Cavaillon Cedex, France
[b] Laboratoire de Structure et Propriétés de l'Etat Solide, UMR CNRS 8008,
Université des Sciences et Technologies de Lille,
59655 Villeneuve d'Ascq Cedex, France
[c] Arrhenius Laboratory, University of Stockholm, 10691 Stockholm, Sweden

**Abstract**

Spark plasma sintering (SPS) of MgAl$_2$O$_4$ powder was investigated at temperatures between 1200 and 1300°C. A significant grain growth was observed during densification. The densification rate always exhibits at least one strong minimum, and resumes after an incubation period. Transmission electron microscopy investigations performed on sintered samples never revealed extensive dislocation activity in the elemental grains. The densification mechanism involved during SPS was determined by anisothermal (investigation of the heating stage of a SPS run) and isothermal methods (investigation at given soak temperatures). Grain-boundary sliding, accommodated by an in-series {interface-reaction/lattice diffusion of the O$^{2-}$ anions} mechanism controlled by the interface-reaction step, governs densification. The zero-densification-rate period, detected for all soak temperatures, arise from the difficulty of annealing vacancies, necessary for the densification to proceed. The detection of atomic ledges at grain boundaries and the modification of the stoichiometry of spinel during SPS could be related to the difficulty to anneal vacancies at temperature soaks.

# I. Introduction

Fully dense polycrystalline alumina-magnesia spinel (referred to as spinel hereafter), $MgAl_2O_4$, is an attractive material for its excellent optical properties (in-line transmittance) in the visible to mid-infrared ranges.[1,2] It is currently considered as a cost-effective alternative to monocrystalline sapphire for the manufacturing of infrared-domes, intended to be mounted on the new generation of high speed air-to-air/ground-to-air missiles coming onto the market. However, most of the polycrystalline materials developed up to now exhibit a grain size in the 10–150 µm range, explaining the disappointing mechanical/thermomechanical properties reported.[3,4] A different approach has recently proved possible to obtain fully dense polycrystalline spinel with a grain size well below the micrometer,[5,6] using a sinter/ hot-isostatic pressing (HIP) strategy. a) Address all correspondence to this author.

To simultaneously limit grain growth and obtain nearly maximum densification, spark plasma sintering (SPS) has been successfully applied to other materials, such as TiN,[7] $Al_2O_3$,[8,9] $Si_3N_4$,[10] 3, and 8 mol% yttria-stabilized $ZrO_2$[11,12] and β-SiC.[13] Dense polycrystalline spinel with acceptable optical properties (residual pores are still present in the material) can be processed using SPS.[14] The grain size obtained is nevertheless still in the order of tens of micrometers and should be defined as coarse.[14] We therefore decided to investigate SPS as a densification method for spinel, aiming at combining high densification and grain size below the micrometer.

The sintering behavior of a commercially available spinel powder was investigated for a soak temperature in the range 1200–1300°C, a soak time of 15 min, a heating rate of 100 °C/min, and an applied macroscopic compaction pressure of 25 MPa. The relative density of the sintered samples has been correlated with their average grain size. The mechanism controlling densification during the SPS experiments are investigated and discussed.

| Ca (ppm) | Cr (ppm) | Cu (ppm) | Fe (ppm) | K (ppm) | Na (ppm) | Si (ppm) | Zn (ppm) | S (ppm) | $H_2O$ (wt %) | SSA ($m^2$/g) |
|---|---|---|---|---|---|---|---|---|---|---|
| <10 | <10 | <10 | <10 | 40 | 15 | <10 | <10 | 400 | 0.70 | 30–31 |

TABLE I. Impurities, humidity level, and specific surface area of the raw powder. Note: ppm in weight.

# II. Raw powder and green samples ready for SPS experiments

The commercially available S30CR raw powder (Baikowski Chimie, La Balme de Sillingy, France) was selected as the starting material. The main impurities determined by inductively coupled plasma spectrometry (ICP; Varian Vista Pro, Varian Inc., Palo

Alto, CA) are listed in Table I. Because the raw powder is crystallized from an alum-based process, the residual sulfur concentration appears relatively high (around 400 ppm in weight).

Using Brunauer–Emmett–Teller analysis (BET) measurements (Nova 2000, Quantachrome Instruments, Boynton Beach, FL) the specific surface area (SSA) of the raw powder was found in the range 30–31 m²/g.

Scanning electron microscope (SEM; JSM-6301F, JEOL Ltd., Tokyo, Japan) examinations showed aggregation of the raw powder. The elemental crystallites constituting the aggregates have a spherical shape and an average diameter in the range 55–70 nm, in good agreement with the specific surface area.

X-ray diffraction (XRD; Bruker D5000, Bruker AXS Gmbh, Karlsruhe, Germany) confirmed that only the cubic $MgAl_2O_4$ species (space group 227, $Fd\bar{3}m$, lattice parameter of 8.0831 Å) is present in the raw powder.

High solids loading (53 wt%) water-based slurries were prepared from the S30CR raw powder (ammonium polyacrylate was incorporated as a dispersant). After deagglomeration, optimal blending, and degassing, samples were slip-casted in porous plaster molds.

Once setting was completed, green samples were left in a drying oven for a few hours. After a debinding step in air (480°C/3 h), samples were ready for the SPS experiments. The diameter of the samples was typically 19.7 mm and their thickness 6 mm (the diameter/thickness ratio was always above 2.5, which strongly minimized the axial density gradient during the SPS experiments). For all samples, the relative green density was around 42% (the theoretical density for spinel has been calculated to be 3.579 g/cc from the elemental lattice). The fracture surface in Fig. 1 shows the typical microstructure observed in green samples.

### III. Experimental Procedure

All the tests were conducted in vacuum, with the slipcasted debinded samples, on equipment (SPS-2080, SPS Syntex Inc., Kanagawa, Japan) located at the Arrhenius Laboratory (Stockholm University, Sweden).

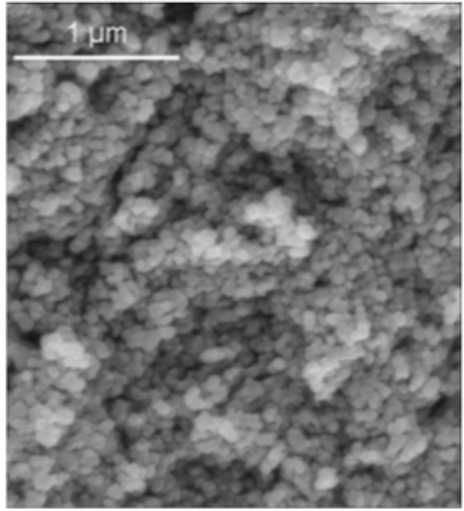

FIG. 1. Typical aspect of the green microstructure obtained after slip casting—fracture surface.

For each test, a graphite die (internal diameter of 20 mm, thickness of 15 mm) was filled with one green sample and mounted on the SPS equipment (graphite punches). A heating rate of 100°C/min, a macroscopic compaction pressure of 25 MPa (applied at room temperature), and the standard 12:2 pulse sequence for the direct current (dc)[8] were chosen.

The temperature was obtained from an optical pyrometer focused on the outer surface of the graphite die. This temperature was therefore not the one actually seen by the powder. Modeling of the temperature distribution during field-assisted sintering, performed on a TZ3Y raw powder, has been recently conducted.[15] For graphite die geometry similar to that used here, the temperature difference between the specimen center (4.25 mm thick) and the external pyrometer focused on the outer die wall surface was calculated to be around 80°C.[15]

The thermal conductivity of spinel is much higher than that of TZ3Y (respectively 14.6 and 2.5 W/m/K). The maximum temperature difference between the outer surface of the die and the powder compact should therefore be around 80°C (probably less) during the SPS experiments performed on spinel samples.

During all tests (heating, soak, and cooling), the height variation of the powder bed ($\Delta L = L - L_0 < 0$, L is the instantaneous height and $L_0$ the initial height of the powder bed when the macroscopic pressure is applied at room temperature) was precisely measured. Each test was corrected to account for the dimensional variations of the SPS equipment (a blank test with a fully dense polycrystalline spinel sample positioned in the die was performed and then subtracted to the test result). The instantaneous sample height variation and the relative density D are linked by the following relationship:

$$D = \left(\frac{L_f}{L}\right)D_f \quad \text{(eq. 1)}$$

where $L_f$ is the final height, L the instantaneous height, and $D_f$ the final relative density.

The apparent density of the sintered samples was measured using the Archimedes method with deionized water (three measurements were made for each sample). The final relative density, $D_f$, was obtained using a theoretical density of 3.579 g/cc for stoichiometric spinel.

Thin foils were prepared from the central zone of as sintered samples by slicing and mechanical polishing, followed by ion milling. The foils were covered with a thin layer of graphite and observed using a Philips CM30 microscope (Philips Research Laboratories, Eindhoven, The Netherlands, acceleration voltage of 300 kV, point-to-point resolution of 0.19 nm) equipped with an energy dispersive spectroscopy (EDS) microanalysis system (Thermo Electron Corporation, Waltham, MA, Noran system equipped with an ultrathin window). The general microstructure was observed in bright field mode. Local EDS analyses were performed on one sample using the scanning transmission electron microscopy (STEM) mode in the center of the elemental grains (10 measurements), at grain boundaries (10 measurements) and at triple points (9 measurements) using a probe size of 5.6 nm. Quantitative analyses were carried out using the Doukhan–Van Cappellen method, based on electroneutrality of the specimen to access the local thin foil thickness.[16] Additional investigations were also performed using the high-resolution transmission electron microscopy (HRTEM) mode at grain boundaries.

TEM was also used to evaluate the grain size for each sintered sample. A line-intercept method taking into account at least 150 grains (with a three-dimensional correction factor determined to be 1.2, approximating the grains to spheres[17]) was used.

The densification rate variation (relative to temperature), 1/D dD/dT, is shown as a function of temperature up to 1400°C in Fig. 2. Densification starts around 750°C and three regimes are observed. (i) From 750 to 1150°C the densification rate (relative to temperature) increases continuously, (ii) between 1150 and 1315°C the densification rate (relative to temperature) has a constant value around $2\times10^{-3}$/°C, and (iii) above 1315°C the densification rate (relative to temperature) increases abruptly.

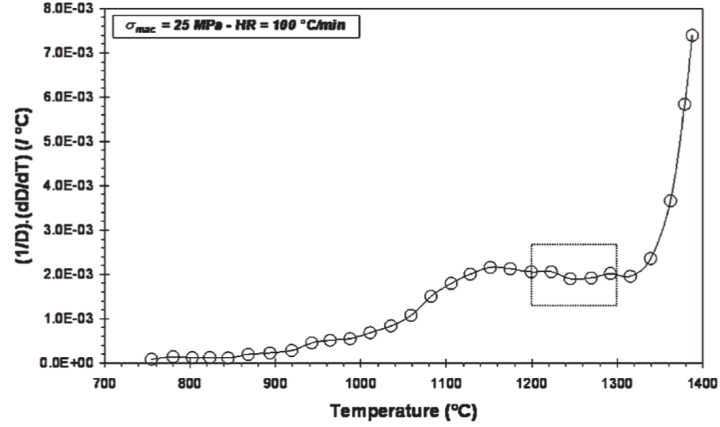

FIG. 2. Densification rate (relative to temperature) curve obtained on a slip casted sample. The heating rate is fixed to 100°C/min, the applied macroscopic compaction pressure to 25 MPa (the compaction pressure is applied at room temperature). The rectangle determines the temperature range where subsequent SPS experiments are performed.

For this study, the 1200 to 1300°C temperature range, where the densification rate (relative to temperature) is always around $2\times10^{-3}$/°C, was selected. Even if the real temperature in the compact is not precisely known [finite element modeling (FEM) calculations are required] we assume that the temperature difference between the matrix, where the pyrometer is focalized, and the powder bed is at maximum 80°C and constant in the range 1200–1300°C. The following SPS conditions in vacuum were then selected: (i) soak temperature = 1200–1225–1250–1275–1300°C; (ii) soak time = 15 min; (iii) heating rate (HR) = 100°C/min; (iv) macroscopic compaction stress = 25 MPa (applied at room temperature); and (v) 12:2 pulse configuration.

In the past, mass transport during sintering, with or without an external load, has been considered close to that occurring in high temperature creep.[18,19] Assuming an approach similar to Mukherjee for the creep of dense metals,[20] it has been proposed that the SPS kinetic equation can be written as[11,21]:

$$\frac{1}{\mu_{eff}}\frac{1}{D}\frac{dD}{dt} = K\frac{e^{\frac{Q_d}{RT}}}{T}\left(\frac{b}{G}\right)^p\left(\frac{\sigma_{eff}}{\mu_{eff}}\right)^n \qquad \text{(eq. 2)}$$

where D is the instantaneous relative density of the compact, t the time, $\mu_{eff}$ the instantaneous shear modulus of the compact, K a constant, R the gas constant, T the absolute temperature, $Q_d$ the apparent activation energy of the mechanism controlling densification, **b** the Burgers vector (close to the lattice parameter), G the grain size, and $\sigma_{eff}$ the instantaneous effective stress acting on the compact. It was also proposed that $\mu_{eff}$, and $\sigma_{eff}$ can be written as[11,21]:

$$\mu_{eff} = \frac{E_{th}}{2(1+v_{eff})} \frac{D-D_0}{1-D_0} \qquad \text{(eq. 3)}$$

$$\sigma_{eff} = \frac{1-D_0}{D^2(1-D_0)} \sigma_{mac} \qquad \text{(eq. 4)}$$

where $E_{th}$ is the Young's modulus of the theoretically dense $MgAl_2O_4$ material (its variation as a function of temperature can be found in Ref. 22), $v_{eff}$ the effective Poisson's ratio (a value of 0.26 is chosen for all experimental conditions), $D_0$ the starting green density of the powder compact (42%), and $\sigma_{mac}$ the macroscopic compaction pressure (25 MPa).

Equation (4), proposed for the effective stress, approximates the individual grains to spheres, independently of the relative density value. To be perfectly correct, the real effective stress should also incorporate an additional term related to a pressureless kind of driving force, which becomes prominent when the relative density approaches 1. An intuitive expression could be $2\gamma/r$, the same as for pressureless sintering, where g is the surface energy and r the pore radius. It is also possible that this pressureless kind of driving force is counterbalanced by another force related to gas pressure that develops within closing pores during the SPS runs. More investigations (theoretical and experimental) are needed to develop a more precise expression for the effective stress. In the meantime, the macroscopic compaction pressure alone is assumed to be responsible for the driving force operating during the SPS experiments we completed.

From Eq. (2), following the procedure suggested by Brook,[18] $Q_d$, p, and n can be determined. These are the key parameters enabling the identification of the mechanisms controlling densification of the powder bed during the SPS experiments.

## IV. Results

Figure 3 shows the densification curves at soak. The higher the temperature, the higher the relative density is after sintering. The shape of the curves is nevertheless uncommon and different to that reported for TZ3Y.[11] The densification curves for all temperatures, but 1200°C, adopt a wavy shape (the temperature regulation of the SPS equipment is good enough to ensure it has no influence on the curves shape). Some "plateaus," where the relative density remains almost constant during a certain period of time, are observed before densification resumes. This behavior is very similar to that observed during hot pressing of spinel powders.[23] It is also close to that observed during creep experiments on fully dense polycrystalline spinel.[24,25]

The variations of 1/D dD/dt as a function of D for all the test temperatures are shown in Fig. 4. When SPS was performed between 1225 and 1300°C, strong minima where densification of the compact is strongly reduced are detected. This corresponds to the "plateau" periods observed in Fig. 3. After the minima, 1/D dD/dt increased again and finally collapsed at the end of the test. When the SPS temperature is 1200°C, the same behavior is observed, though clearly less pronounced. It is also interesting to note that the minima move to higher relative densities with an increase in the test temperature. To understand this behavior, it could be interesting to perform interrupted SPS experiments (with a high cooling rate to "freeze" the structure) and observe the resulting microstructure by TEM.

The grain size versus relative density trajectory, referred to as sintering path, and obtained with results from the different samples sintered by SPS is shown in Fig. 5. It is compared to the sintering path obtained for the same material (slip-casted samples) densified using pressureless sintering (PS) in air (the heating rate is only 10°C/min, in comparison to 100°C/min for the SPS experiments) at 1500 °C. In the green samples, the elemental crystallite size is around 55 to 70 nm (Fig. 1). A significant grain growth is observed during both the SPS and PS runs. The sintering path of the SPS material is below the one of the PS material. For a given relative density, the final grain size obtained using SPS will be smaller. A good approximation for the grain size/relative density trajectory obtained by SPS is given by the dashed line visible in Fig. 5. Subsequently, for the rest of the work, the following expression linking G to D will be used:

$$G = \alpha e^{\beta D} \quad \text{(eq. 5)}$$

with α = 0.0018 and β = 5.3586.

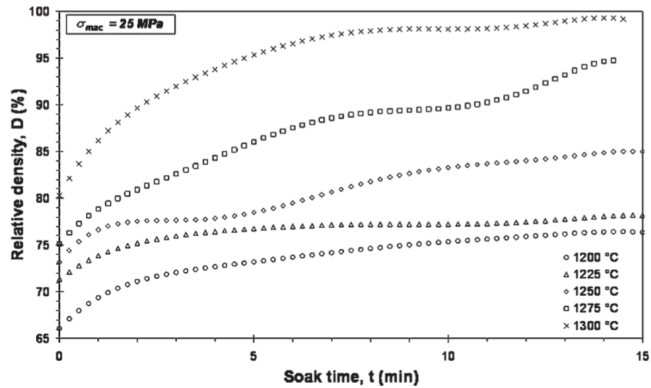

FIG. 3. Densification curves obtained for different SPS temperatures as a function of soak time. The applied macroscopic compaction pressure is fixed to 25 MPa.

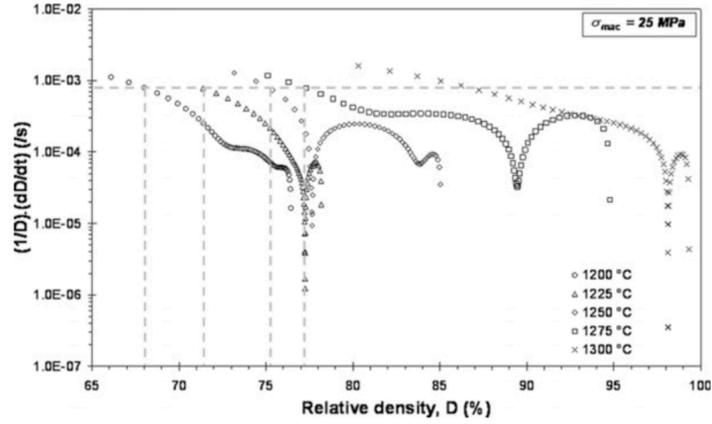

FIG. 4. Densification rate (relative to time) versus relative density for the different soak temperatures. The applied macroscopic compaction pressure is fixed to 25 MPa and the soak time to 15 min.

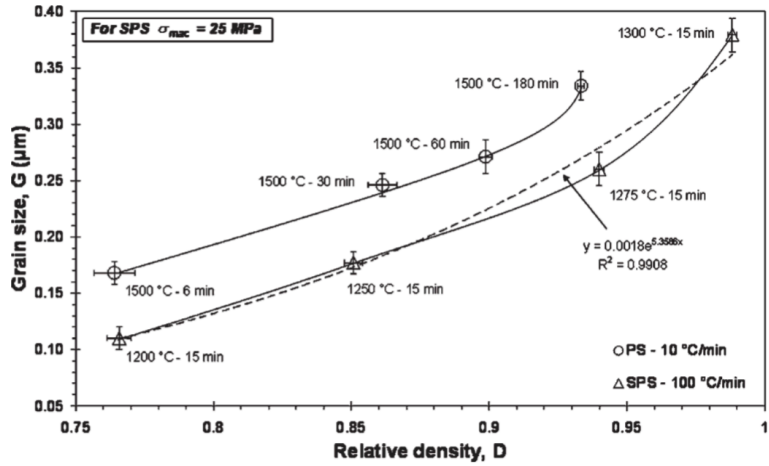

FIG. 5. Sintering path for the SPS material (heating rate of 100°C/min, applied macroscopic compaction pressure of 25 MPa), after 15 min at the different soak temperatures. It is compared to the one obtained from pressureless sintered runs (PS, heating rate of 10°C/min), performed on the same kind of green samples and for soak times up to 180 min.

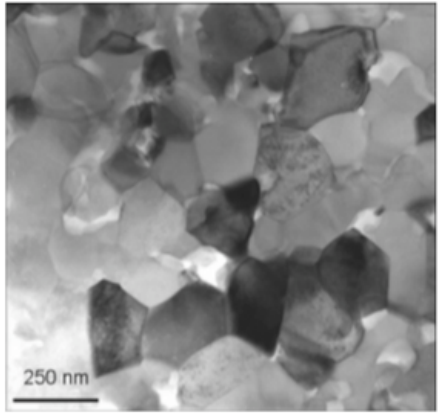

FIG. 6. Typical aspect of the microstructure of the sample obtained by SPS at 1250°C during 15 min. The heating rate is fixed to 100 °C/min and the applied macroscopic compaction pressure to 25 MPa.

Even if the residual porosity level is different, the typical microstructures for samples sintered at 1200°C (76.58% relative density, 110-nm grain size), 1250°C (85.08% relative density, 177-nm grain size), and 1275°C (93.99% relative density, 260-nm grain size) for 15 min are qualitatively similar. As an example, the microstructure observed on the sample sintered at 1250°C is shown in Fig. 6. For all samples (i) the residual porosity is located at grain boundaries (intergranular porosity), and is homogeneous in size (no aggregation to form large pores). Its spatial distribution is also homogeneous; (ii) the spinel grains constituting the matrix exhibit an equiaxed shape and a narrow grain size distribution; and (iii) no dislocation activity was detected in the elemental grains or at grain boundaries (multi two-axis tilting of the thin foils).

The typical microstructure developed in the sample sintered at 1300°C for 15 min (98.82% relative density, 379-nm grain size) is shown in Figs. 7(a) and 7(b). Most of the porosity has disappeared, only a few residual pores are observed at triple points and homogenously distributed throughout the thin foil. Some intragranular pores are detected [see white arrow in Fig. 7(a)], though their total content is low. The grains still exhibit an equiaxed shape but it seems that the average grain size from one elemental crystal to another one is more variable, as compared to previous samples. Some grains also exhibit an intragranular dislocation activity [Fig. 7(b)]. The slip systems activated were not investigated (use of the weak-beam method, determination of the g vector). Such dislocations are nevertheless clearly homogeneously detected in the sample sintered at 1300°C and not in the other ones sintered at a lower temperature. However, despite the clear presence of such events, the general dislocation activity is quantitatively considered as low. At that time, it is not clear why a dislocation activity is only observed for a SPS temperature of 1300°C. It could be interesting in the future to investigate SPS temperatures above 1300°C to analyze if dislocation motion becomes a prominent phenomenon that could have an influence on the control of densification.

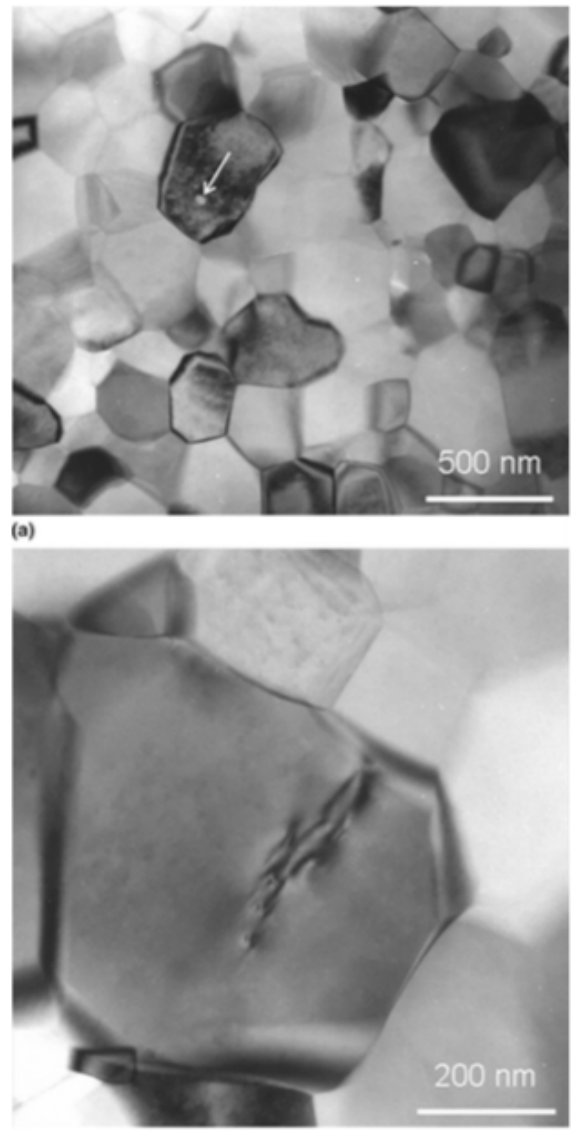

FIG. 7. Typical aspect of the microstructure of the sample obtained by SPS at 1300°C during 15 min. The heating rate is fixed to 100 °C/ min and the applied macroscopic compaction pressure to 25 MPa. (a) No residual intergranular pores are detected but few entrapped intragranular pores are observed (arrow), (b) some grains contain an intragranular dislocation activity.

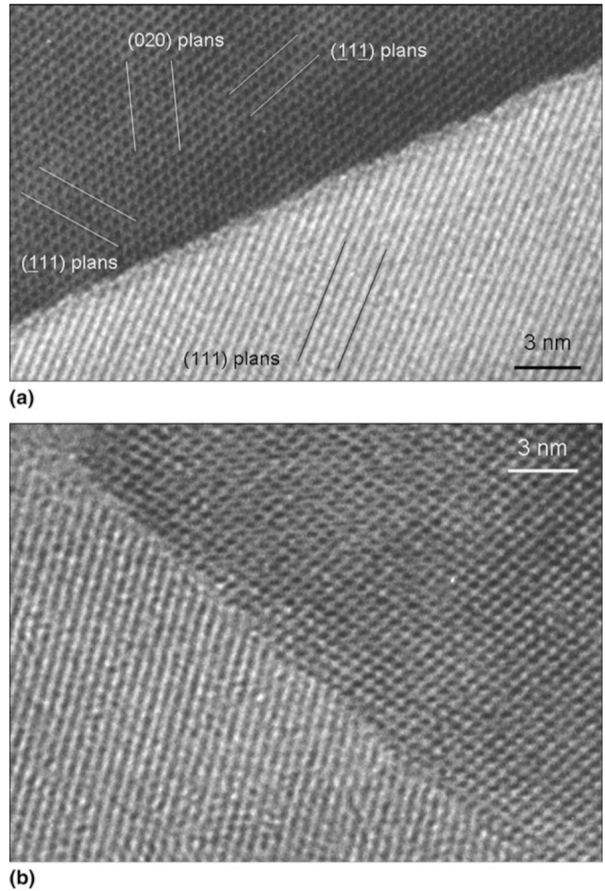

FIG. 8. Typical aspect of grain boundaries observed by HRTEM in the sample obtained by SPS at 1300°C during 15 min. The heating rate is fixed to 100°C/min and the applied macroscopic compaction pressure to 25 MPa. (a) No residual amorphous thin film is detected, (b) observations out of focus show that the grain boundaries are not perfectly straight and atomic ledges are present at most of them.

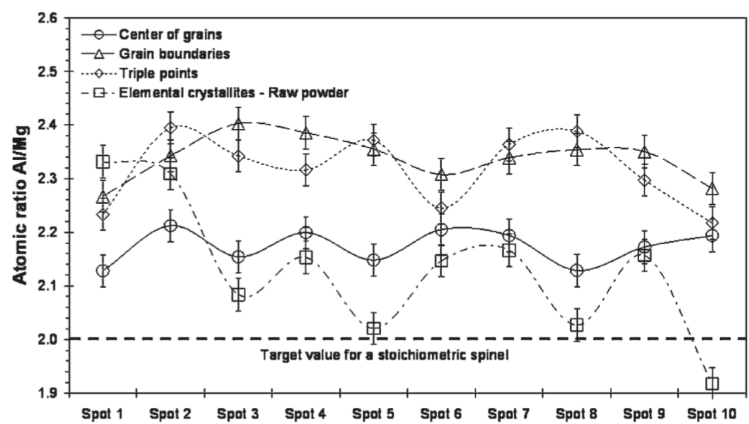

FIG. 9. EDS nanoanalysis performed by STEM in the sample obtained by SPS at 1300°C during 15 min. The heating rate is fixed to 100°C/min and the applied macroscopic compaction pressure to 25 MPa. For all zones analyzed (center of grains, grain boundaries, triple points), the results show that the dense polycrystalline material is not a stoichiometric spinel anymore.

For all sintering temperatures, densification during SPS of the spinel samples investigated is not controlled by a mechanism involving dislocations (gliding or climbing) contribution. In fact, the microstructures observed are in good agreement with a densification mechanism based on a grain-boundary sliding/diffusion accommodated process.

Figure 8(a) shows the typical aspect of grain boundaries observed using HRTEM. Grain boundaries appear depleted of any amorphous thin film. Using higher magnification, with a nonoptimal focus, most of the grain boundaries do not appear perfectly flat and exhibit atomic ledges, as shown in Fig. 8(b).

EDS nanoanalyses (Fig. 9) have been performed for the sample sintered at 1300°C for 15 min (98.82% relative density, 379-nm grain size). For all locations (center of grain, grain boundary, triple point), the atomic Al/Mg ratio is clearly above 2 (standard deviation is in all cases +/-0.03). For comparison, the O/Al ratio is constant, with an average value of 1.96+/-0.01 for the grain centers versus 1.93+/- 0.01 for grain boundaries and triple points. The following chemical compositions have been determined: (i) $Mg_{0.922}Al_2O_{3.922}$ for grain centers; (ii) $Mg_{0.855}Al_2O_{3.855}$ for grain boundaries; and (iii) $Mg_{0.862}Al_2O_{3.862}$ for triple points.

Using similar experimental conditions, EDS nanoanalyses have been conducted on the elemental crystallites constituting the raw powder. Results are also shown in Fig. 9. Therefore, the average composition $Mg_{0.939}Al_2O_{3.939}$ for crystallite centers has been determined. The O/Al ratio is constant for all crystallites with an average value of 1.98 0.02.

Clearly, there is a change of stoichiometry during SPS of spinel and an impoverishment in MgO is observed, especially at grain boundaries and at triple points. However, at that time, the modification of stoichiometry was not understood (critical temperature, critical relative density, pressure effect, etc). This change of stoichiometry might also have an influence on the space charge at grain boundaries and, consequently, on the diffusion process that controls densification and grain growth. We will come back later on to that point in Sec. V. Finally, it is also possible that the slight brown color of the sintered samples does matter with the astoichiometry finally obtained. But carbon contamination, from a CO containing residual atmosphere during the SPS experiments, cannot be excluded (graphite heating die; a rotary pump is generating vacuum in which the residual gases can give rise to an atmosphere containing $CO_2$ possibly transforming to CO at high pressure within the shrinking pores). If such a contamination occurs during the SPS tests, the expression proposed for the densification rate should be modified accordingly (in that case the driving force is not the effective pressure alone anymore, a contribution from gas pressure developing within closed pores has to be subtracted). Complementary investigations are needed to clarify this point.

## V. Discussion

To discriminate the mechanisms controlling the densification of the spinel powder during SPS, it is necessary to determine the values of the $Q_d$, p, and n parameters in Eq. (2). Rearranging Eq. (2) yields:

$$Ln\left[\frac{T}{\mu_{eff}}\frac{1}{D}\frac{dD}{dT}\frac{dT}{dt}\right] = -\frac{Q_d}{RT} - pLn(G) + nLn\left(\frac{\sigma_{eff}}{\mu_{eff}}\right) + K' \text{ (eq. 6)}$$

where G is given by relation (5) and dT/dt is the heating rate during the SPS experiment.

Phenomenological models have been developed to describe high-temperature creep behavior for ceramic polycrystals.[26,27] Such models can be adapted to an SPS problematic, where the densification mechanism is not based on dislocations activity, as it is the case for the runs performed on spinel (see TEM observations reported in Sec. IV).

If the grain boundaries are perfect sources/sinks of vacancies, the n and p parameters in relations (2) and (6) can have the following values[26]: (i) n = 1, p = 2: the densification mechanism is grain-boundary sliding accommodated by volume diffusion and the apparent activation energy has a bulk character; and (ii) n = 1, p = 3: the densification mechanism is grain-boundary sliding accommodated by grain-boundary diffusion and the apparent activation energy has a grain-boundary character.

If the grain boundaries are not perfect sources/sinks of vacancies, the n and p parameters in relations (2) and (6) have the following values[27]: (i) n = 2, p = 1: the densification mechanism is grain-boundary sliding accommodated by an in-series {interface-reaction/lattice diffusion} mechanism controlled by the interface-reaction step and the apparent activation energy has a bulk character; and (ii) n = 2, p = 2: the densification mechanism is grain boundary sliding accommodated by an in-series {interface reaction/grain-boundary diffusion} mechanism controlled by the interface-reaction step and the apparent activation energy has a grain-boundary character.

The heating part of an SPS run, for temperatures between 940 and 1300°C, can also be exploited. Imposing a given value for the activation energy Q, it is possible, using the Excel Solver function (Microsoft Excel, Microsoft France, Courtaboeuf, France), to calculate the corresponding p, n, and $K_0$ parameters that enable the left side of relation (6) to be equal to its right side by minimization of the residual sum of squares (RSS).

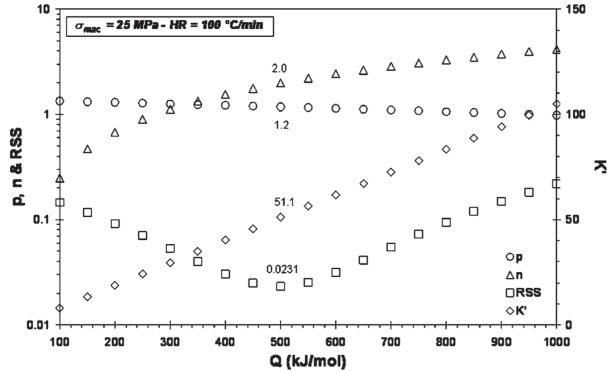

FIG. 10. Determination of the densification mechanism using an anisothermal method. The heating rate is fixed to 100°C/min and the applied macroscopic compaction pressure to 25 MPa. The heating portion of a SPS experiment is used, activation energy values are imposed (Q) and the corresponding p, n, and $K_0$ parameters involved in relation (6) are calculated using Excel Solver function. Best result is for the lowest RSS (residual sum of squares).

The p, n, $K_0$, and RSS values, obtained for each value of Q imposed, are summarized in Fig. 10. Clearly, the minimum RSS value is obtained when n and p have a value of 2.0 ±0.1 and 1.2±0.0, respectively. In that case, $K_0$ is 51.1±2.1 and $Q_d$ has a value of 500±20 kJ/ mol. Because of the obtained values for n (around 2) and p (close to 1), the apparent activation energy has a bulk character. In the past, it has been shown that the activation energy for oxygen self-diffusion in monocrystalline stoichiometric spinel is in the range 415–500 kJ/mol,[28–30] comparable to the value of 500 kJ/mol obtained there. For comparison, the activation energies for the selfdiffusion of $Mg^{2+}$ cations, in the same kind of spinel monocrystal, has been determined to be around 200 kJ/mol.[31] No value was found in the literature for the selfdiffusion of the $Al^{3+}$ cations, although it can be indirectly estimated. Martinelli et al.[32] concluded from conductivity experiments that, at 1000°C, magnesium is the more mobile cation. Independently, a $Mg^{2+} \leftrightarrow Al^{3+}$ interdiffusion activation energy of 235 kJ/mol was determined in by Watson.[33] Assuming that migration of the aluminum cation is the rate limiting step in this process, as suggested by Martinelli, then Murphy et al.[34] concluded that 235 kJ/mol could be a good evaluation for the activation energy for $Al^{3+}$ self-diffusion in spinel. Both values reported for $Mg^{2+}$ and $Al^{3+}$ cations are much lower than the apparent activation energy for densification obtained from these SPS experiments on spinel.

We therefore propose that grain-boundary sliding, accommodated by an in-series {interface-reaction/lattice diffusion of the $O_2$ anions} mechanism controlled by the interface reaction step, governs densification of our spinel samples during the heating portion of the SPS experiments we performed (at least between 940 and 1300°C).

Regarding the densification curves obtained for the different soak temperatures, combining relations (2) and (5) yields:

$$\frac{1}{\mu_{eff}}\frac{1}{D}\frac{dD}{dt} = K_0 \frac{e^{-\frac{Q_d}{RT}}}{T} e^{-\beta p D} \left(\frac{\sigma_{eff}}{\mu_{eff}}\right)^n \text{ (eq. 7)}$$

where $K_0$ is a constant.

Assuming only one constant value of $Q_d$ (coherency with the results obtained using the anisothermal method, no reason for a change in densification mechanism between the heating portion and the early stages of the soaks), the slope of the straight line obtained when plotting

$$Ln\left(\frac{1}{\mu_{eff}}\frac{1}{D}\frac{dD}{dt}\right) + \beta p D = f\left[Ln\left(\frac{\sigma_{eff}}{\mu_{eff}}\right)\right]$$

corresponds to the n value. Knowing the n value, the slope of the straight line obtained when plotting

$$Ln\left(\frac{T}{\mu_{eff}}\left(\frac{\sigma_{eff}}{\mu_{eff}}\right)^n \frac{1}{D}\frac{dD}{dt}\right) + \beta p D = f\left[\frac{1}{T}\right]$$

Using a fixed value of 1 for p, Fig. 11 shows the variations of
$$Ln\left(\frac{1}{\mu_{eff}}\frac{1}{D}\frac{dD}{dt}\right) + \beta p D$$
as a function of $Ln(\sigma_{eff}/\mu_{eff})$ for the different soak temperatures selected. It seems that n exhibits an average value of 2 when densification progresses at the beginning of the soaks, for temperatures between 1200 and 1275°C, in good agreement with the ideal combination (n = 2, p = 1) determined from the heating portion of an SPS experiment (see above). But in all cases the

$$Ln\left(\frac{1}{\mu_{eff}}\frac{1}{D}\frac{dD}{dt}\right) + \beta p D = f\left[Ln\left(\frac{\sigma_{eff}}{\mu_{eff}}\right)\right]$$

trajectories differ from a straight line after some period of time at soak. When the soak temperature is fixed to 1300°C, the apparent value of n is lower than 2.

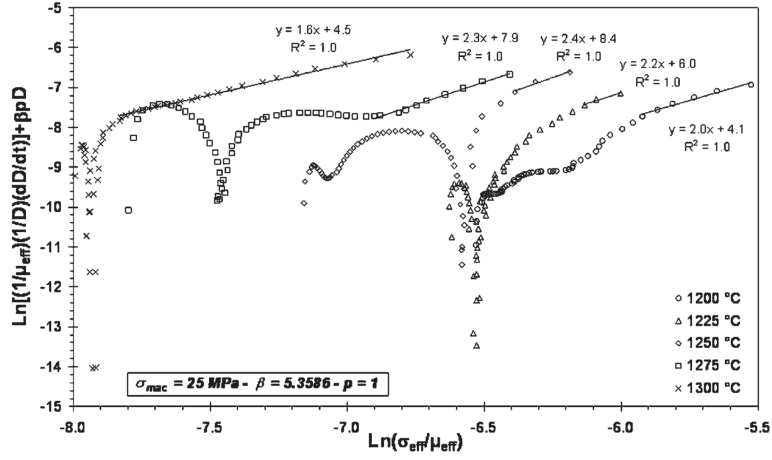

FIG. 11. Initial effective stress exponent calculated with relation (11) for the different soak temperatures. The heating rate is fixed to 100°C/min and the applied macroscopic compaction pressure to 25 MPa. The stress exponent values are the slopes of the different straight lines.

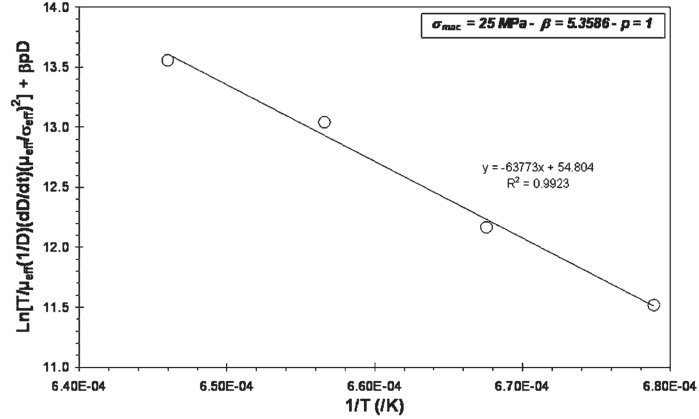

FIG. 12. Apparent activation energy for densification, $Q_d$, when n≈2. The slope of the straight line is $Q_d/R$. The heating rate is fixed to 100°C/min and the applied macroscopic compaction pressure to 25 MPa.

Figure 4 presents the formalism used to calculate $Q_d$ with a stress exponent of 2. A fixed value of 1/D dD/dt ($8\times10^{-4}$/s) has been chosen to fall within the regime where n has a value of 2, for all the soak temperatures investigated. The corresponding relative density values are then accessible (dashed lines, Fig. 4), which allow us to calculate the corresponding values of $\sigma_{eff}$ and $\mu_{eff}$. Finally the variation of

$$Ln\left(\frac{T}{\mu_{eff}}\left(\frac{\mu_{eff}}{\sigma_{eff}}\right)^2 \frac{1}{D}\frac{dD}{dt}\right) + \beta pD$$

as a function of 1/T is plotted (Fig. 12). According to relation (10), a value of $Q_d$ around 530±30 kJ/mol is calculated (XLSTAT solver, Addinsoft France, Paris, France).

This value is similar to what has been obtained using the heating portion of an SPS experiment (500±20 kJ/mol). Therefore, it is proposed that grain-boundary sliding, accommodated by an in-series {interface-reaction/lattice diffusion of the O2- anions} mechanism controlled by the interface-reaction step, is still governing densification of our spinel samples at the beginning of the soak, for SPS temperatures in the range 1200–1275°C. For soak temperatures of 1200, 1225, and 1250°C, the period where n is close to 2 is followed by an abrupt densification hardening regime (the instantaneous value of n increases continuously), corresponding to a strong decrease of the instantaneous relative densification rate (Fig. 4). A similar trend is observed for a soak temperature of 1300°C, where n has a value between 1 and 2. When the soak temperature is 1275°C, the densification hardening regime is preceded by a period where n is close to 0. After the hardening period, densification resumes, for all soak temperatures. This corresponds to a densification softening period.

Such densification or strain hardening/densification or strain softening behavior has already been observed during hot pressing of spinel powder[24] and high temperature creep experiments on polycrystalline spinel.[25] The softening and hardening were related to a change in the internal stresses, depending on a decrease and increase in the density of the intragranular dislocations, respectively, whose motions contribute to the relaxation of stress concentrations exerted through the predominant mechanism of grain-boundary sliding. It was proposed that densification/deformation was controlled by the continuous recovery of the dislocations, limited by lattice diffusion of the oxygen ions.

Clearly, this hypothesis is not in agreement with the typical microstructures observed for the different samples obtained here by SPS. For all temperatures, no extensive dislocation activity has been reported in the elemental grains constituting the sintered samples (see Sec. IV). It is proposed here that the densification hardening, corresponding to a zero-densification-rate period, is originating from the difficulty to anneal vacancies, which is the driving force for the densification to proceed. At each soak temperature, after a certain period of time, vacancies are accumulating because the annealing step stops. Then densification also stops. An incubation time is then necessary to anneal enough vacancies to resume densification. Atomic ledges have been detected at grain boundaries using HRTEM [Fig. 8(b)]. The EDS nanoanalyses have also shown that the stoichiometry of spinel is changing during SPS (Fig. 9). Both events, at the nanoscale, could be related to the difficulty to anneal vacancies at the soak temperatures.

The astoichiometry is amplified at grain boundaries and at triple points, in comparison to the center of the grains. The EDS analyses suggest an excess of Mg and O vacancies in these areas. If the concentrations for each kind of vacancy are not the same, the consequence will be an excess of negative positive charge at grain boundaries and triple points, depending on which concentration is the highest. Therefore, the residual electrical charge will be compensated by an opposite electrical charge cloud in the surrounding grains, at the close vicinity of the grain boundaries/triple points, called space charge.[35] The space-charge region creates an electric potential and therefore modifies the conditions of the charged defect formation and diffusion that can explain the perturbation of the interface-reaction (zero densification-rate period), claimed to control densification during our SPS experiments. Interesting works have been published on the stoichiometry variation of polycrystalline spinel at grain boundaries.[36,37] No successes have been reported for the characterization of the space charge region alone in such materials. In fact, such investigations are difficult, since the space-charge dimension could be lower than the lowest spot size available on the best TEM/STEM equipments. The fact that the SPS technology also involves the submission of the compact to an electric field has perhaps also an influence on the properties of a possible space-charge region and consequently on the possible perturbation of the interface-reaction. To investigate such a last effect, the same kind of spinel samples will be densified using standard hot pressing in the future.

Another comment may be added, regarding the determination of n, p, and $Q_d$. The method investigating the heating part of a SPS run is fairly straightforward. Inversely, more questionable are the results obtained when investigating the densification curves at soak. In such case, it may be difficult to determine a precise value for n because the hardening phenomenon appears rapidly at soak (Fig. 11, 1225 and 1250°C). At least an approximated value may be extrapolated using few points at the beginning of the soak.

## VI. Conclusions

SPS of a stoichiometric alumina-magnesia spinel powder, shaped by slip casting, has been investigated in vacuum, in the 1200 to 1300°C temperature range. The other experimental parameters were a heating rate of 100°C/min, an applied macroscopic compaction pressure of 25 MPa, a soak time of 15 min, and the use of the standard 12:2 pulse configuration.

For all soak temperatures, grain growth is significant during densification. Accounting for this phenomenon, the mechanism controlling the densification of spinel powder during the SPS experiments has been identified. It is proposed that grain-boundary sliding, accommodated by an in-series {interface-reaction/lattice diffusion of the $O^{2-}$ anions} mechanism controlled by the interface reaction step, governs densification. This hypothesis is in good agreement with microstructural observations, performed on the SPS samples, using TEM. For each soak temperature, a zero-densification-rate period is observed. In our case, the lack of dislocation activity in the elemental grains after SPS implies that the zero-densification-rate period is related to the stop of the interface reaction that controls the annealing of vacancies. More investigations are now required to identify precisely the mechanisms responsible for such a zero-densification-rate period.